\documentclass[aps,pre,preprint]{revtex4-1}

\usepackage{graphicx}
\usepackage{amsmath,amssymb}
\usepackage{psfrag}
\usepackage{color}

\usepackage{tikz}
\usetikzlibrary{shapes.geometric}

\begin{document}

\newcommand{\rhoc}{\rho_{\scriptscriptstyle \rm cross}}
\newcommand{\kB}{k_{\scriptscriptstyle \rm B}}

\date{\today}

%\title{Paradoxical cooperative transport}
%\title{Exclusion processes with selective kinetic constraints}
\title{Cooperative transport with selective kinetic constraints}

\author{Mauro Sellitto} 
\email{mauro.sellitto@unicampania.it, mauro.sellitto@gmail.com}
\affiliation{Dipartimento di Ingegneria, Universit\`a della Campania
  ``Luigi Vanvitelli'', Via Roma 29, I-81031 Aversa (CE), Italy.
\\ The Abdus Salam International Centre for Theoretical Physics,
Strada Costiera~11, I-34151 Trieste, Italy.  }

\begin{abstract}
We introduce and study a family of cooperative exclusion
processes whose microscopic dynamics is governed by selective kinetic
constraints. They display, in sharp contrast to the simple
symmetric exclusion process, density profiles that can be concave,
convex or both, depending on the density of boundary particle
reservoirs. A mean-field analysis based on a diffusion equation with a
density-dependent diffusion coefficient qualitatively reproduces this
behaviour, and suggests its occurrence in liquids with a diffusivity
anomaly.
\end{abstract}

\maketitle

{\bf Introduction.} -- Cooperative transport underlies the variety of
complex behaviours observed in soft condensed matter systems where the
subtle interplay of weak entropic forces and nonequilibrium fluxes
conspire to sustain highly flexible organized structures and their
biological functionality~\cite{Nelson}.  While a first-principle
characterization of these often counterintuituive features remains
difficult, coarse grained approaches based on exclusion
processes~\cite{nerve,acid}, have been successfully applied to several
problems such as polymerization kinetics on nucleic acid templates,
molecular motors, and cellular transport, to name only a
few~\cite{Pavel,Andreas,Kirone}. Moreover, and perhaps more
importantly, they have also come to play a paradigmatic role in recent
advances of nonequilibrium statistical
mechanics~\cite{Kirone,Derrida}.

Constrained exclusion processes, in which particle hopping requires a
{\em minimum} number of vacant neighbours~\cite{KoAn, RiSo}, are
already known to display rich nonequilibrium properties such as
non-Fickian transport, differential negative resistance, heterogeneous
dynamics, non-standard fluctuation relation, extending regimes of
anomalous diffusion, dynamical free energy
singularities~\cite{Se_ratchet,Se_acep,Se_FT,TuPi,TuPiSe}, and
systematic approaches for computing transport diffusion coefficient
with increasing accuracy have been recently developed for this class
of non-gradient stochastic lattice gases~\cite{Teomy,Arita}.

In this paper we investigate a more general family of cooperative
exclusion processes in which the number of vacant neighbours required
for hopping is not determined by a minimum threshold but rather can be
any specific set of non-negative integers (lower than lattice
coordination number). This idea of {\em selective} kinetic constraints
has been recently introduced in the context of bootstrap percolation
to provide models of multiple hybrid phase transitions in a fully
homogeneous environment~\cite{Se_SBP}.  We are primarily interested
here in the emergence of complex convexity-change density profiles and
dynamical effective particle attraction or repulsion generated by the
subtle interplay of nonequilibrium fluxes and kinetic constraints, in
the absence of any static interaction (apart from hard-core
exclusion).

Our results show that constraining some transition probabilities to
zero leads to density profiles that are globally not bounded by those
obtained in the absence of constraints with the same boundary
condition. As this property cannot be obviously anticipated and may
appear rather paradoxical at a glance we first discuss its occurrence
in the macroscopic context of a diffusion equation with a
density-dependent diffusion coefficient, where it can be easily
understood as a consequence of a diffusivity anomaly. Then, we see how
it can be realized microscopically on a lattice and, finally, present
numerical results showing that the hydrodynamic behaviour of
cooperative exclusion processes is well accounted for, at least
qualitatively, by a diffusion equation in which the diffusion
coefficient is naively estimated by neglecting particle correlations.

\bigskip

{\bf Diffusion equation.} -- Let us assume that the transport process
occurs in a interval of size $L$ and the system dynamics is governed
by the partial differential equation:
\begin{eqnarray}
  \frac{\partial \rho}{\partial t} = \frac{\partial}{\partial x}
  \left[ D(\rho) \frac{\partial \rho}{\partial x} \right],
\end{eqnarray}
with boundary condition $\rho(0,t)=\rho_0$ and $\rho(L,t)=\rho_1$,
where $\rho (x,t)$ is the particle density at position $x\in [0,L]$
and time $t$, and $D(\rho)$ is the density-dependent diffusion
coefficient.  In the steady state the particle current:
\begin{eqnarray}
  J = - D(\rho) \frac{\partial \rho}{\partial x} 
\end{eqnarray}
is uniform and constant, and the particle density profile is
implicitly determined by:
\begin{eqnarray}
  x(\rho) = a + b \int D(\rho) d \rho  ,
  \label{eq:x}
\end{eqnarray}
where the constants $a$ and $b$ are fixed by the boundary condition
$x(\rho_0)=0,\, x(\rho_1)=L$.  When the diffusion coefficient does not
depend on particle density the steady state current $J$ is
proportional to $\rho_1 - \rho_0$, and the steady state density
profile is linear:
\begin{eqnarray}
  \rho_{\scriptscriptstyle \rm linear} (x) = \rho_0 + (\rho_1 -
  \rho_0) \frac{x}{L}.
\end{eqnarray}
On a lattice this situation is realised by the simple
symmetric exclusion process ({\small SSEP}), in which particles
interact only with hard core exclusion.
When the diffusion coefficient depends on $\rho(x)$ something more
interesting happens. For systems with slow dynamics, in which
$D(\rho)$ monotonically decreases at high particle density, one
generally observes convex density profiles, $\rho_{\scriptscriptstyle
  \rm slow}''(x)<0$, leading to the overall upper bound:
\begin{eqnarray}
  \rho_{\scriptscriptstyle \rm slow}(x) < \rho_{\scriptscriptstyle \rm
    linear} (x), \quad \forall x \in (0,L).
  \label{inequality}
\end{eqnarray}
A simple analytically solvable instance is given by a power-law
density-dependent diffusion coefficient and a lattice realisation is
provided by {\em constrained} exclusion processes~\cite{Se_ratchet}.
Although the relation $x'(\rho) \propto D(\rho)$ guarantees that
$\rho(x)$ is always monotonic in $x$, the possibility of
convexity-change profiles may still occur for more general forms of
cooperative dynamics, and precisely in systems with a diffusivity
anomaly in which relaxation dynamics becomes faster upon isothermal
compression. This is the opposite of what happens in a ``normal''
liquid, where a reduction in the specific volume generally limits the
ability of molecules to move. Such an unusual feature is shared by a
number of liquids, typically with directional interactions (the most
important of which being certainly water), and is manifested by a
$D(\rho)$ that exhibits a minimum or a maximum in some density range
(see Refs.~\cite{Francesco,Sal,Ludo,Massimo} for some examples).
Evidently, one observes a convex (concave) $\rho(x)$ when $D'(\rho)<0$
($D'(\rho)>0$). In particular, there will be some boundary densities,
for which $\rho(x)$ will change convexity at location
$x(\rho^{\star})$ such that $D'(\rho^{\star})=0$.  In this case, no
global inequality like Eq.~\eqref{inequality} can be stated.
The key question is to identify the conditions on microscopic
transition probabilities under which such unusual features arise in
the absence of static interactions.

\begin{figure}
  \centering 
  \includegraphics[scale=.46]{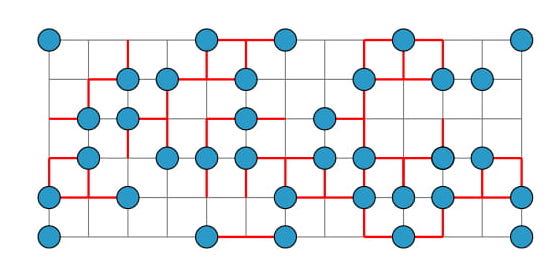}
  \caption{Example of lattice configuration in a $2D$ cooperative
    exclusion process in which particle moves to a vacant neighbour
    are allowed only if the particle does not have two vacant
    neighbours, before and after the move. Forbidden particle hopping
    are denoted by red lattice bonds (full periodic boundary condition
    is assumed).}
  \label{fig.forbidden_moves}
\end{figure}

\bigskip

{\bf Microscopic models.} -- We argue that the family of cooperative
exclusion processes ({\small CEP}) we now introduce features
convexity-change density profiles when kinetic constraints are of
selective type. Selective here means that particle hopping requires
some specific numbers of vacant neighours before and after the move,
rather than a minimum threshold $m$, as in usual kinetically
constrained models (KCM)~\cite{RiSo}.
When constraints act selectively in an intermediate range of density,
or also at both low and high densities, we expect that the diffusion
coefficient is not monotonic (i.e., it exhibits a minimum or a
maximum, respectively).
One can give a useful mean-field estimation for {\small CEP} diffusion
coefficient by neglecting correlations between nearby particles, in
terms of the probability of not having certain numbers of vacant
neighbours before and after the move, just as done for {\small
  KCM}~\cite{TuPiSe,Teomy,Arita}:
\begin{eqnarray}
  D^{\scriptscriptstyle \rm NC}_{\scriptscriptstyle {\mathcal
      S}}(\rho) & = & \left[ 1 - \sum_{i \, \in \, {\mathcal S} } {c-1
      \choose i} \rho^{c-1-i} (1-\rho)^i \right]^2,
  \label{eq:D_NC}
\end{eqnarray}
where $c$ is the lattice coordination number and ${\mathcal S}$ is the
non-negative integer sequence representing the numbers of vacant
neighbours (not counting the departure and target site, before and
after the move, respectively) for which a particle move is {\em not}
allowed.
In the above no-correlation ({\small NC}) approximation, binomial
terms account for the multiplicity of possible configurations of
particles and vacancies, around the departure and target site, that
prevent hopping, and the power 2 comes from the detailed balance
condition.
When the integer sequence ${\mathcal S}$ is gapless, i.e., ${\mathcal
  S}=\{0, 1,2, \dots, m-2\}$, the usual {\em cumulative} form of
kinetic constraints is recovered (in which a move is allowed only when
the particle has at least $m - 1$ vacant sites, not counting the
departure and target site, before and after the move, respectively),
and the diffusion coefficient decreases monotonically all the way down
to $\rho=1$.  When there are gaps in ${\mathcal S}$, e.g., when the sum
over ${\mathcal S}$ in Eq.~\eqref{eq:D_NC} includes only terms in
$\rho^j (1-\rho)^k$ (with nonzero $j$ and $k$), or both terms
$\rho^{c-1}$ and $(1-\rho)^{c-1}$, the diffusion coefficient may
exhibit extrema in an intermediate range of density. In analogy with
its cumulative counterpart, and consistently with Ref.~\cite{Arita},
we expect that Eq.~\eqref{eq:D_NC} yields a general upper bound for
the actual diffusion coefficient, which can be obtained to the lowest
order approximation from a variational principle due to
Varadhan~\cite{Spohn}.

\bigskip

\begin{figure}
  \centering
  \includegraphics[scale=.6]{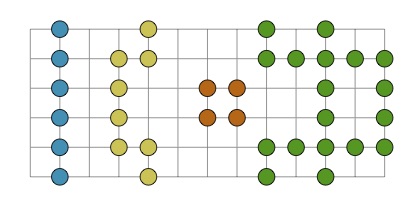}
  \caption{Four examples of metastable particle clusters (each with a
    different colour) that are permanently frozen as long as they
    remain isolated (periodic boundary condition in the vertical
    direction).  The selective constraint allows for a particle move
    only if the particle does not have two vacant neighbours, before
    and after the move.}
\label{fig.metastable_configurations}
\end{figure}

{\bf Numerical results.} -- To substantiate our claim more concretely
we now consider an exclusion process on a two dimensional square
lattice in which a particle move to a nearby empty site occurs
only when the particle does not have two vacant neighbours, before and
after the move. In the previously introduced classification scheme
(which does not take into account the departure and target sites) this
corresponds to ${\mathcal S}=\{1\}$. A representative example of
disallowed moves is shown by red bonds in the particle configuration
of Fig.~\ref{fig.forbidden_moves}.
Dynamics with the above selective constraint obeys detailed balance
and equilibrium measure is trivial just as in any other KCM of glassy
dynamics~\cite{KoAn,RiSo}.  In the limit of very low and very high
particle density the selective kinetic constraint plays no role and
the standard {\small SSEP} is recovered.  In the intermediate range of
density dynamics becomes weakly cooperative. In fact, particle
rearrangements occur rather quickly and, although there exist
particle configurations which are permanently frozen as long as they
remain completely isolated (see
Fig.~\ref{fig.metastable_configurations} for some examples), their
lifetime is actually quite short: as soon as a single particle comes
close to their border the metastable structure breaks up relatively
faster. This leads to a dynamics that is ergodic at any density. We
have explicitly checked, for an equilibrium system with fixed particle
number, that the mean-square displacement is normal after a very short
transient and no anomalous diffusion is observed on extended time
scales.
\begin{figure}[htbp]
  \includegraphics[scale=.6]{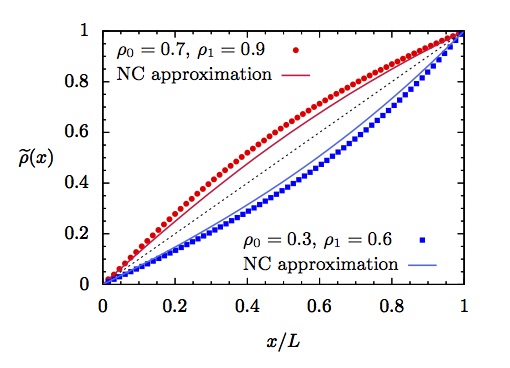}
  \caption{Normalised density profiles. Square and circle symbols
    denote numerical simulations data, solid lines correspond to
    no-correlation ({\small NC}) approximation. The straight dotted
    line is the {\small SSEP} reference profile.}
\label{fig.rho_03_06}
\end{figure}

\bigskip

We have performed Monte Carlo simulation of the nonequilibrium steady
state attained by the system when it is coupled to particle reservoirs
at its edges. The system consists of a square lattice of size $L
\times 2L$ with periodic boundary conditions in all directions in
order to avoid spurious edge effects due to the kinetic constraint.
This amounts to running two independent bulk systems sharing the same
reservoirs, one located at $z = L$ (reservoir with density $\rho_1$)
and the other at $z=0$ (reservoir with density $\rho_0$).  Numerical
results for local density profiles are averaged over $10^6$
independent configurations and have been compared with the inverse
function of density profile:
\begin{eqnarray}
  x(\rho) & = & a + b \left( \rho -2 \rho^3 + \frac{3}{2} \rho^4 +
  \frac{9}{5} \rho^5 - 3 \rho^6 + \frac{9}{7} \rho^7 \right)
  \label{eq:x_rho}
\end{eqnarray}
(with constants $a$ and $b$ fixed by boundary condition),
%$x(\rho_0)=0,\, x(\rho_1)=L$), 
which is obtained exactly from Eq.~\eqref{eq:x} in the no-correlation
approximation for the diffusion coefficient:
\begin{eqnarray}
  D^{\scriptscriptstyle \rm NC}_{\scriptscriptstyle 1}(\rho) & = &
  \left[ 1 - 3 \, \rho^2 \, (1- \rho) \right]^2 .
  \label{eq:D_1}
\end{eqnarray}
This function has a minimum at $\rho^{\star}=2/3$.  According to our
analysis of diffusion equation, this means that one should observe
globally convex or concave profiles depending on whether 
$\rho_0$ and $\rho_1$ are both smaller or larger than $\rho^{\star}$,
respectively. Or also, a density profile that changes convexity at
position $x(\rho^{\star})$ when $\rho_0 < \rho^{\star}<\rho_1$.
For the sake of simplicity we use the normalised density profile
$\widetilde{\rho}(x)$ defined as:
\begin{eqnarray}
  \widetilde{\rho}(x) \equiv \frac{\rho(x) - \rho_0}{\rho_1-\rho_0}.
\end{eqnarray}
Figure~\ref{fig.rho_03_06} shows the numerical results for
$\rho_0=0.3,\,\rho_1=0.6$ and $\rho_0=0.7,\,\rho_1=0.9$ for a system
of linear size $L = 128$ (the absence of finite-size effects has been
checked with $L = 64$ and $L = 256$).  In contrast with the
boundary-driven {\small KA} model~\cite{Se_ratchet,Teomy}, we observe
convex profiles for large values of reservoirs density, and concave
ones when $\rho_0,\, \rho_1$ are both smaller than $\rho^{\star}$. The
comparison with analytical predictions of the diffusion equation obtained
in the {\small NC} approximation (full lines in
Fig.~\ref{fig.rho_03_06}) shows that discrepancies are rather mild, as
also observed in Ref.~\cite{Teomy} for the {\small KA} model.

\begin{figure}[htbp]
  \includegraphics[scale=.6]{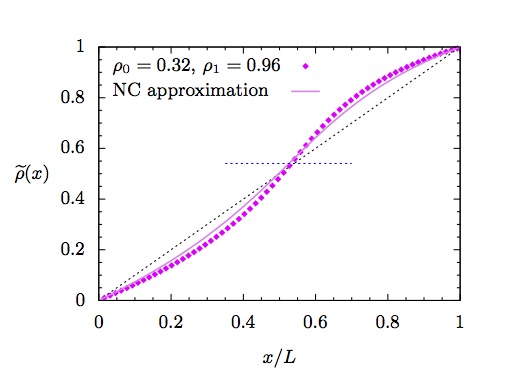}
  \caption{Convex-to-concave crossover normalised density profiles.
    Symbols denote numerical simulations data, the solid line correspond to
    the no-correlation ({\small NC})) approximation. The horizontal
    segment locates the estimated $\rho^{\star}=2/3$ (corresponding to
    $\widetilde{\rho}^{\star}\simeq 0.54...)$ in the {\small NC}
    approximation. (The straight dotted line is the reference {\small
      SSEP} profile).}
\label{fig.rho_032_096}
\end{figure}
\begin{figure}[htbp]
  \includegraphics[scale=.6]{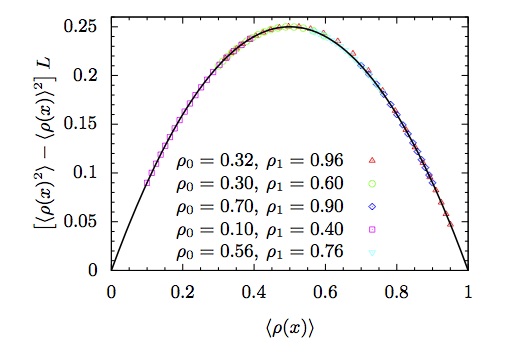}
  \caption{Parametric plot of transverse density fluctuations vs
    density profile for several values of reservoir density, $\rho_0$
    and $\rho_1$, and linear system size $L=128$. The solid line is
    the result expected from uncorrelated {\small SSEP}-like
    transverse fluctuations.}
\label{fig.drho}
\end{figure}

Figure~\ref{fig.rho_032_096} shows a convexity-change density profile
obtained for $\rho_0 < \rho^{\star} < \rho_1$, that is
$\rho_0=0.32,\,\rho_1=0.96$. Also in this case we see that predictions
are well confirmed, in particular the position at which convexity
changes.  A more refined approximation scheme can be certainly
obtained by exploiting the systematic approaches recently developed in
Refs.~\cite{Teomy,Arita}.  We leave this to future works. What we
would like to emphasize here is the rather unusual nature of density
profiles we have obtained.
Even though, from a microscopic point of view, there is evidently no
{\em static} force, the overall effect of disallowing some moves and
imposing a current flux in the system favours those kinetic paths that
bring particles closer or farther, as they travel from the high to the
low density reservoirs.  This suggests that the interplay of
nonequilibrium drive and kinetic constraints leads to the emergence of
effective dynamical interactions, which may be attractive or repulsive
depending on the sign of $D'(\rho)$ and are manifested in convex,
concave or convexity-change density profiles.
Perhaps surprisingly, {\em transverse} local density fluctuations are
not affected by the shape of density profiles and turns out to be
always uncorrelated (for sufficiently large system size), irrispective
of the boundary driving force, just as in {\small SSEP}:
\begin{eqnarray}
  L \left[ \langle \rho(x)^2 \rangle - \langle \rho(x) \rangle^2
    \right] = \langle \rho(x)\rangle (1 - \langle \rho(x) \rangle).
\end{eqnarray}
This is shown in the parametric plot of Fig.~\ref{fig.drho}.

We have carried out the same analysis for another cooperative
exclusion process on a square lattice, in which a particle move is
allowed only when the particle does not have one or four vacant
neighbours, before and after the move. In our classification scheme
this is represented by the selective constraint ${\mathcal S}
=\{0,\,3\}$, which corresponds to the particle-hole symmetry invariant
analog of the {\small $2D$} {\small KA} model~\cite{KoAn}.
On a Bethe lattice it has two glass transition critical points (at low
and high density, related by the particle-hole symmetry). On a square
lattice an argument similar to that developed in~\cite{ToBiFi} ensures
that dynamics is ergodic at any density, even though it is strongly
cooperative and sluggish when $\rho \to 0$ or $\rho \to 1$ (diffusion
becoming singular near these points). The {\small NC} approximation for
the diffusion coefficient in this case is:
\begin{eqnarray}
  D^{\scriptscriptstyle \rm NC}_{\scriptscriptstyle 0,3 }(\rho) & = &
  \left[ 1 - \rho^3 - (1- \rho)^3 \right]^2 ,
  \label{eq.D03}
\end{eqnarray}
which has a maximum at $\rho^{\star}=1/2$. This gives profile shapes
that are just the opposite of those discussed previously for the case
${\mathcal S} =\{1\}$, i.e., concavity for $\rho_0 <1/2,\,\rho_1
<1/2$, convexity for $\rho_0 >1/2,\,\rho_1 >1/2$, and convexity-change
profiles for $\rho_0 <1/2<\rho_1$. The comparison with Monte Carlo
results shows that these predictions are well reproduced with small
discrepancies and improving accuracy as $\rho_0$ and $\rho_1$
increasingly depart from 0 and 1.  Thus, we conclude that the
properties discussed above are quite general of the {\small CEP}
family and, in spite of substantial particle correlation arising from
cooperative dynamics, the hydrodynamic behaviour is well captured, at
least qualitative, by the nonlinear diffusion equation. More exotic
features, like density profiles with multiple changes of convexity,
can also be observed provided that the system has a larger lattice
coordination number and extra gaps in ${\mathcal S}$, as suggested by
Eq.~\eqref{eq:D_NC}, but we have not explored this further.

\bigskip

{\bf Conclusions} -- In summary, we have shown that cooperative
exclusion processes with selective kinetic constraints and trivial
equilibrium properties, display density profiles with a considerable
rich structure in the absence of any static interaction. Such features
are the macroscopic nonequilibrium manifestation of attractive and
repulsive statistical forces that arise from microscopically
constrained current-carrying nonequilibrium steady states. This is
nicely captured by a diffusion equation with an approximated
density-dependent diffusion coefficient that neglects particle
correlations.  Quite independently of the nature of kinetic
constraints, which must be considered as a coarse-grained description
of the real microscopic dynamics, the analysis of the diffusion
equation shows that convexity-change profiles are intimately related
to a non-monotonic density-dependent diffusion coefficient.
Therefore, liquids with a diffusivity anomaly and soft matter systems
with a reentrant dynamics~\cite{Francesco,Sal,Ludo,Massimo} are the
most suitable candidates in which the above predictions can be tested.


\begin{thebibliography}{35}

\bibitem{Nelson} P. Nelson, {\em Biological Physics} (New York: WH
  Freeman, 2004).

\bibitem{nerve} A.L. Hodgkin and R. D. Keynes, {\em The Journal of
  Physiology} {\bf 128}, 61 (1955).
  
\bibitem{acid} J.T. MacDonald, J.H. Gibbs and A. Pipkin, {\em
  Biopolymers} {\bf 6}, 1 (1968).

\bibitem{Pavel} P.L. Krapivsky, S. Redner, and E. Ben-Naim, {\em A
  Kinetic View of Statistical Physics}, (Cambridge: Cambridge
  University Press, 2010).

\bibitem{Andreas} A. Schadschneider, D. Chowdhury, and K. Nishinari,
  {\em Stochastic Transport in Complex Systems: From Molecules to
    Vehicles} (Amsterdam: Elsevier, 2011).

\bibitem{Kirone} T. Chou, K. Mallick and R.K.P. Zia,
  {\em Rep. Prog. Phys.} {\bf 74}, 116601 (2011).

\bibitem{Derrida} B. Derrida, {\em J. Stat. Mech.} P07023
  (2007).

\bibitem{KoAn} W. Kob and H.C. Andersen, Phys. Rev. E {\bf 48}, 4364
  (1993). 

\bibitem{RiSo} For a review, see: F. Ritort and P. Sollich, {\it
  Adv. Phys.} {\bf 52}, 219 (2003).

\bibitem{Se_ratchet} M. Sellitto, {\em Phys. Rev. E} {\bf 65},
  020101(R) (2002).

\bibitem{Se_acep} M. Sellitto, {\em Phys. Rev.  Lett.} {\bf 101},
  048301 (2008).

\bibitem{Se_FT} M. Sellitto, {\em Phys. Rev.  E} {\bf 80}, 011134
  (2009).

\bibitem{TuPi} F. Turci and E. Pitard, {\em Europhys. Lett.} {\bf 94},
  10003 (2011).
 
\bibitem{TuPiSe} F. Turci, E. Pitard, and M. Sellitto, {\em
  Phys. Rev. E} {\bf 86}, 031112 (2012).

\bibitem{Teomy} E. Teomy and Y. Shokef, {\em Phys. Rev. E} {\bf 95},
  022124 (2017).

\bibitem{Arita} C. Arita, P.L. Krapivsky, and K. Mallick, {\em
  J. Phys. A: Math. Theor.} {\bf 51}, 125002 (2018).

\bibitem{Se_SBP} M. Sellitto, to appear on {\em J. Stat. Mech.}
  (2019).

\bibitem{Francesco} F.W. Starr, F. Sciortino, and H.E. Stanley,
% Dynamics of simulated water under pressure, 
Phys. Rev. E {\bf 60}, 6757 (1999).

\bibitem{Sal} M.C. Rechtsman, F.H. Stillinger, and S. Torquato,
% Negative Thermal Expansion in Single-Component Systems with
% Isotropic Interactions,
J. Phys. Chem. A {\bf 111}, 12816 (2007).

\bibitem{Ludo} L. Berthier, A.J. Moreno, and G. Szamel, 
%Increasing the density melts ultrasoft colloidal glasses,
Phys. Rev. E {\bf 82}, 060501(R) (2010).

\bibitem{Massimo} M. Pica Ciamarra and P. Sollich, 
%The first jamming crossover: Geometric and mechanical features,
J. Chem. Phys. {\bf 138}, 12A529 (2013).

\bibitem{Spohn} H. Spohn, {\em Large Scale Dynamics of Interacting
  Particles} (Heidelberg: Springer-Verlag, 1991).

\bibitem{ToBiFi} C. Toninelli, G. Biroli, and D.S. Fisher, {\em
  J. Stat. Phys.} {\bf 120}, 167 (2005).

\end{thebibliography}
\end{document}